\documentclass{elsart5p}
\usepackage{amssymb,graphicx}
\begin{document}
\begin{frontmatter}
\title{Momentum-resolved charge excitations in high-$T_c$ cuprates\\
studied by resonant inelastic x-ray scattering}
\author[jaea]{K. Ishii\corauthref{cor1}}
\corauth[cor1]{Corresponding author.}
\ead{kenji@spring8.or.jp}
\author[jaea,esrf]{M. Hoesch}
\author[jaea]{T. Inami}
\author[jaea]{K. Kuzushita}
\author[jaea]{K. Ohwada}
\author[jaea]{M. Tsubota}
\author[jaea,tohoku]{Y. Murakami}
\author[jaea]{J. Mizuki}
\author[jaea,iias]{Y. Endoh}
\author[jaea,imr]{K. Tsutsui}
\author[kyoto,imr]{T. Tohyama}
\author[imr,crest]{S. Maekawa}
\author[imr]{K. Yamada}
\author[osaka,srl]{T. Masui}
\author[osaka,srl]{S. Tajima}
\author[aoyama]{H. Kawashima}
\author[aoyama]{J. Akimitsu}
\address[jaea]{Synchrotron Radiation Research Unit, Japan Atomic Energy
Agency, Hyogo 679-5148, Japan}
\address[esrf]{European Synchrotron Radiation Facility, 38000 Grenoble,
 France}
\address[tohoku]{Department of Physics, Graduate School of Science,
Tohoku University, Sendai 980-8578, Japan}
\address[iias]{International Institute for Advanced Studies, Kizu, Kyoto
619-0025, Japan}
\address[imr]{Institute for Materials Research, Tohoku University, Sendai
980-8577, Japan}
\address[kyoto]{Yukawa Institute for Theoretical Physics, Kyoto
 University, Kyoto 606-8502, Japan}
\address[crest]{CREST, Japan Science and Technology Agency, 4-1-8
Honcho, Kawaguchi 332-0012, Japan}
\address[osaka]{Department of Physics, Graduate School of Science, Osaka
University, Toyonaka 560-0043, Japan}
\address[srl]{Superconducting Research Laboratory, ISTEC, Tokyo 135-0062,
Japan}
\address[aoyama]{Department of Physics and Mathematics, Aoyama-Gakuin
University, Sagamihara, Kanagawa 229-8558, Japan}
\begin{abstract}
We report a Cu $K$-edge resonant inelastic x-ray scattering (RIXS) study
of high-$T_c$ cuprates. Momentum-resolved charge excitations in the
CuO$_2$ plane are examined from parent Mott insulators to carrier-doped
superconductors.  The Mott gap excitation in undoped insulators is found
to commonly show a larger dispersion along the $[\pi,\pi]$ direction
than the $[\pi,0]$ direction. On the other hand, the resonance condition
displays material dependence.  Upon hole doping, the dispersion of the
Mott gap excitation becomes weaker and an intraband excitation appears
as a continuum intensity below the gap at the same time. In the case of
electron doping, the Mott gap excitation is prominent at the zone center
and a dispersive intraband excitation is observed at finite momentum
transfer.
\end{abstract}
\end{frontmatter}

\section{Introduction}
Strongly correlated electron systems have attracted much attention
because they display a wide variety of fascinating physical properties,
such as high-$T_c$ superconductivity in cuprates.  The unveiling of
their electronic structure is mandatory to clarify the mechanisms
underlying these physical behaviors.  Essentially, one can think of the
electronic structure of high-$T_c$ cuprate superconductors as that of a
doped Mott insulator in two dimensions, in which charge and spin degrees
of freedom of the electron are crucial.  Neutron scatting is a powerful
tool to investigate the spin dynamics.  On the other hand, the photon is
a good probe for the charge sector.  Conventional optical methods, such
as optical conductivity measurement, can give information about the
momentum-conserved excitations, but momentum-resolved experiments are
required for our complete understanding of the charge dynamics.  Recent
developments of x-ray sources from synchrotron radiation make this
possible by inelastic x-ray scattering.  Especially, resonant inelastic
x-ray scattering (RIXS) has the advantage of element selectivity and it
enables us to elucidate excitations related to the Cu orbitals by tuning
the incident photon energy to the Cu $K$ absorption edge.  While
angle-resolved photoemission spectroscopy (ARPES) yields one-particle
excitation from the occupied state \cite{Damascelli2003}, RIXS provides
the two-particle excitation spectra, from which one can explore both
occupied and unoccupied states.  In the last decade, the importance of
the RIXS technique has been increasingly recognized for the
investigation of the electronic structure of strongly correlated
electrons systems.


In this paper, we report a comprehensive RIXS study on high-$T_c$
cuprates from parent Mott insulators to carrier-doped superconductors.
We focus on the Mott gap excitation and the intraband excitation below
the gap. The latter emerges when the material is carrier doped.
Evolution of the electronic structure upon carrier doping is a central
issue in the physics of doped Mott insulator and we examine here how
finely it can be probed using RIXS.  Material dependence, including a
peculiar resonance condition in Mott insulators, is discussed as well
through a comparison of our results with previous work.

\section{Mott gap excitation in RIXS}
\begin{figure}[t]
\includegraphics[scale=0.4]{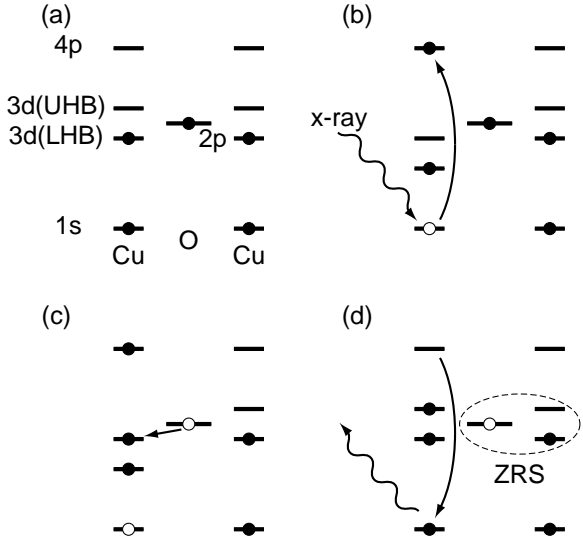}
\caption{Schematic energy diagram of CuO$_2$ plane and RIXS process at
 the Cu $K$-edge.  Two Cu sites and one O $2p_{\sigma}$ level are shown.
 Filled and open circles are electron and hole, respectively.}
\label{fig:rixs}
\end{figure}

In this section, we briefly describe how the Mott gap excitation occurs
in the RIXS process.  Figures\ \ref{fig:rixs} show a schematic energy
diagram of the CuO$_2$ plane and the RIXS process at the Cu
$K$-edge. The parent materials of high-$T_c$ superconductors are
charge-transfer insulators \cite{Zaanen1985}.  The Cu $3d_{x^2-y^2}$
level splits into a lower Hubbard band (LHB) and an upper Hubbard band
(UHB) by strong electron correlation and the O $2p_{\sigma}$ band is
situated between them.  In addition, Cu $1s$ and Cu $4p$ levels, which
participate in the x-ray absorption and emission processes, are included
in the diagram.  Fig.\ \ref{fig:rixs}(a) shows the ground state of a
parent Mott insulator.  X-ray absorption at the left Cu site brings
about a dipole transition from the Cu $1s$ to the $4p$ level [Fig.\
\ref{fig:rixs}(b)].  The $1s$ core hole predominantly scatters a valence
electron.  The energy of the Cu $3d_{x^2-y^2}$ of the left Cu site is
lowered by an attractive Coulomb interaction with the Cu $1s$ core hole.
Then an electron in the O $2p_{\sigma}$ transfers to the UHB to screen
the core-hole potential [Fig.\ref{fig:rixs}(c)].  Finally, the Cu $4p$
electron goes back to the $1s$ level again and a photon is emitted
[Fig.\ \ref{fig:rixs}(d)].  The hole in the O $2p_{\sigma}$ and the
right Cu site form a Zhang-Rice singlet (ZRS) \cite{Zhang1988}.  We call
this scattering event a Mott gap excitation in the CuO$_2$ plane, i.e.,
an excitation from the ZRS band to the UHB.

%

\section{Experimental details}
The RIXS experiments were performed at BL11XU in SPring-8, where a
specially designed inelastic x-ray scattering spectrometer was installed
\cite{Inami2001}.  Using a Si (111) double-crystal monochromator and a
Si (400) channel-cut secondary monochromator, an incident energy
resolution of about 220 meV was obtained.  Horizontally scattered x-rays
were analyzed in energy by a spherically bent Ge (733) crystal.  The
total energy resolution estimated from the full width at half maximum of
the elastic line was about 400 meV.  All spectra were measured at
room temperature.

In Continuation of our previous work on superconductors
\cite{Ishii2005a,Ishii2005b}, their parent Mott insulators,
Nd$_2$CuO$_4$ and YBa$_2$Cu$_3$O$_6$ were measured.  We selected
Ca$_{2-x}$Na$_x$CuO$_2$Br$_2$ as a hole doped sample.  The advantage of
this material in RIXS experiments is the absence of rare earth atoms
that reduces the absorption of x-rays by the sample itself and a larger
scattering intensity can be expected with this advantage.  For the study
of electron doping, we measured Nd$_{2-x}$Ce$_x$CuO$_4$.


Single crystals were prepared for all the samples. The Na concentration
($x$) in Ca$_{2-x}$Na$_x$CuO$_2$Br$_2$ was estimated as $x$ = 0.2 from
the superconducting transition temperature ($T_c$ = 14K)
\cite{Kuroiwa2006}.  YBa$_2$Cu$_3$O$_6$ and
Ca$_{1.8}$Na$_{0.2}$CuO$_2$Br$_2$ were sealed in a beryllium cell filled
with inert gas in order to avoid reaction with oxygen or water in the
air.

The surface of the crystals is perpendicular to the $c$-axis.  The
crystals were mounted with the $c$-axis in the scattering plane.  The
strong two dimensionality of the electronic structure of the CuO$_2$
plane ensures that the momentum dependence along the $c^{\ast}$-axis is
small. Therefore we fixed the $c^{\ast}$ component in the absolute
momentum transfer ($\vec{Q}$) at a value where the scattering angle
($2\theta$) is close to 90$^{\circ}$. This enables us to reduce the
elastic scattering by the polarization factor of the Thomson scattering
\cite{Ishii2005b}.  We measured the momentum dependence of the CuO$_2$
plane in the transverse geometry.  A schematic view of the RIXS
experimental configuration is shown in the inset of Fig.\
\ref{fig:eidep}(d).  The polarization of the incident x-rays
($\vec{\epsilon_i}$) contains nearly equal out-of-plane
($\vec{\epsilon_i} \parallel \vec{c}$) and in-plane ($\vec{\epsilon_i}
\perp \vec{c}$) components.


\section{Results and Discussion}
\subsection{Incident energy dependence and resonance condition of the
  Mott insulators}
\begin{figure*}[phtb]
\includegraphics[scale=0.34]{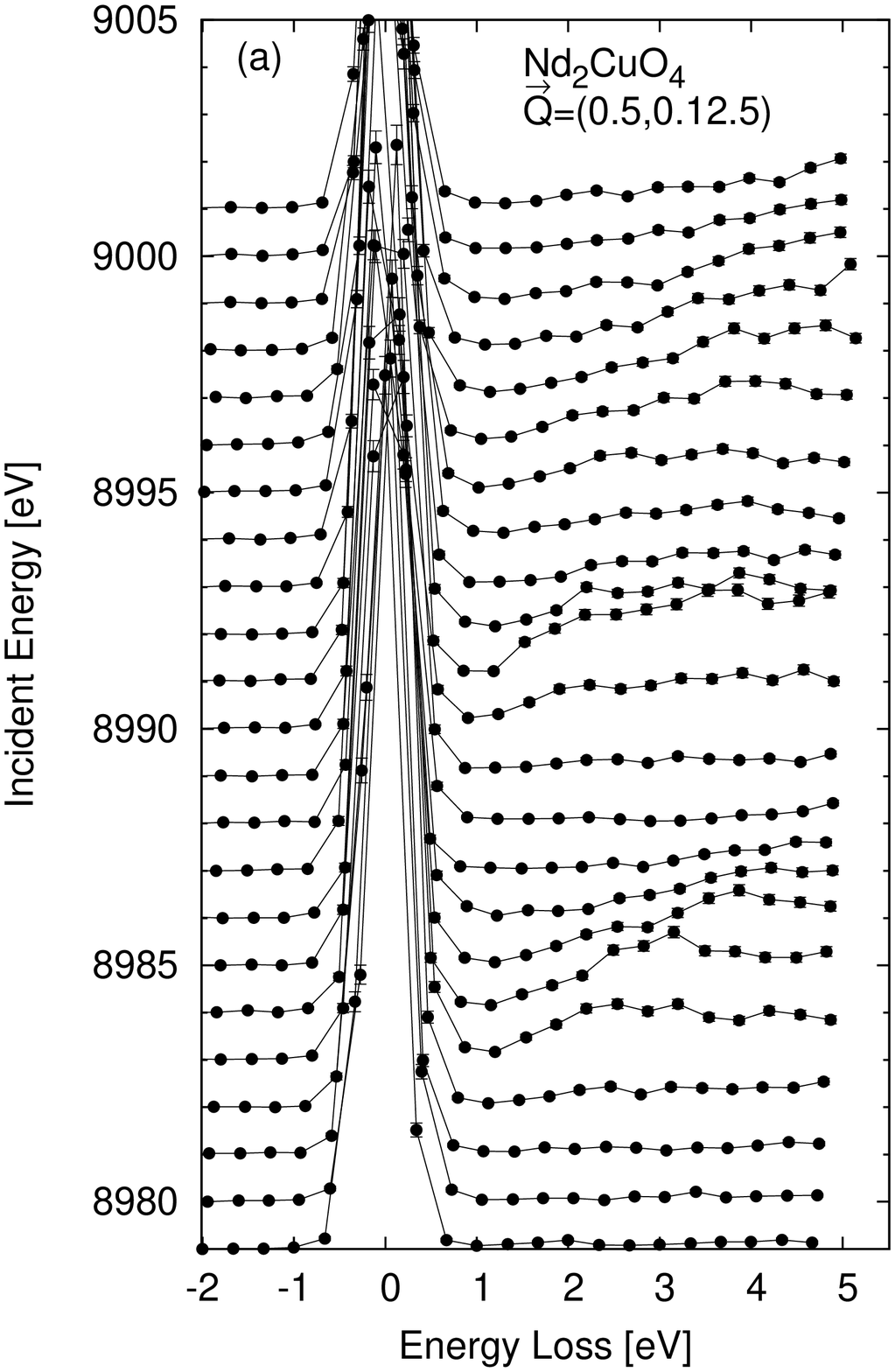}
\includegraphics[scale=0.34]{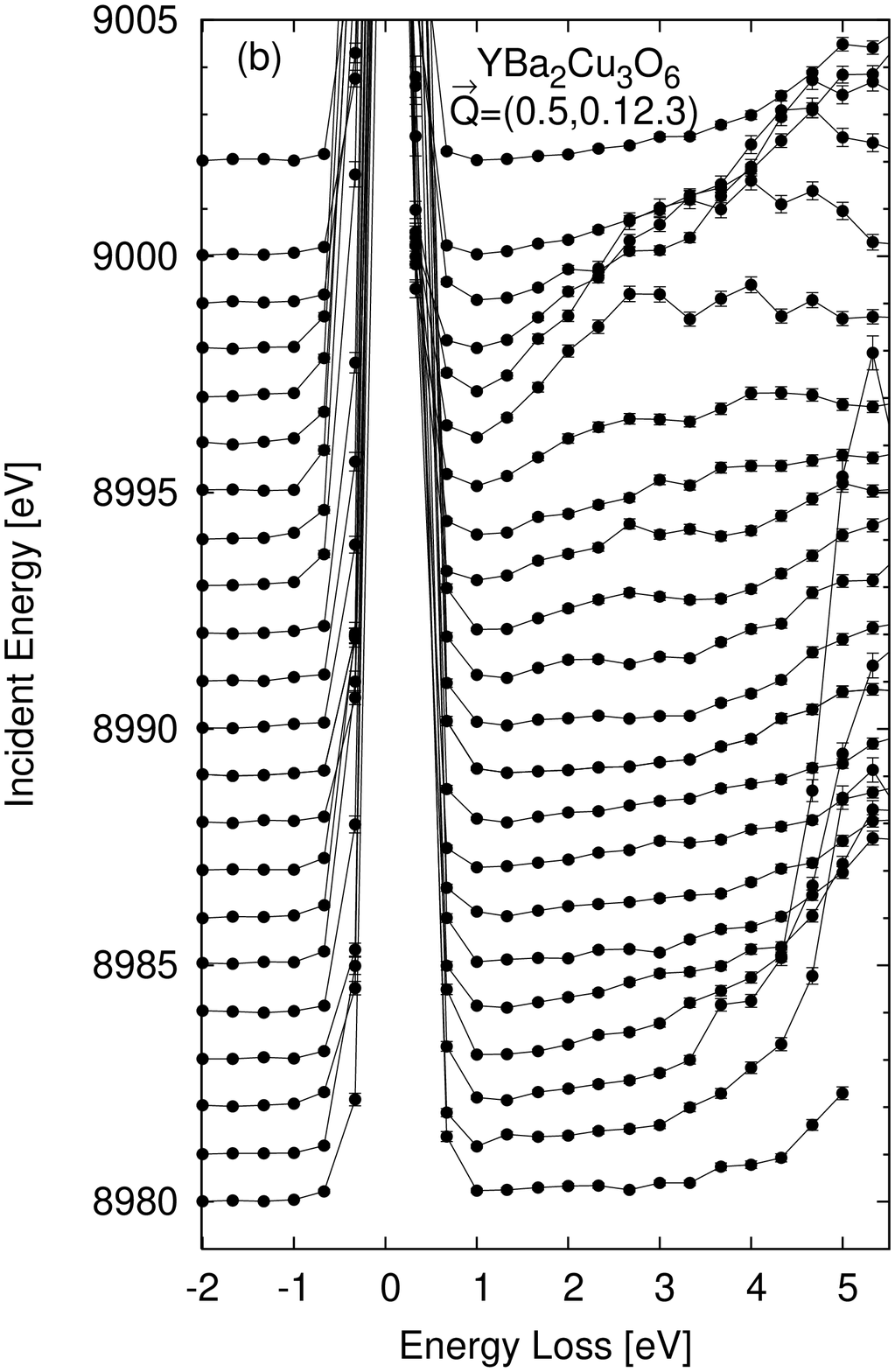}
\includegraphics[scale=0.34]{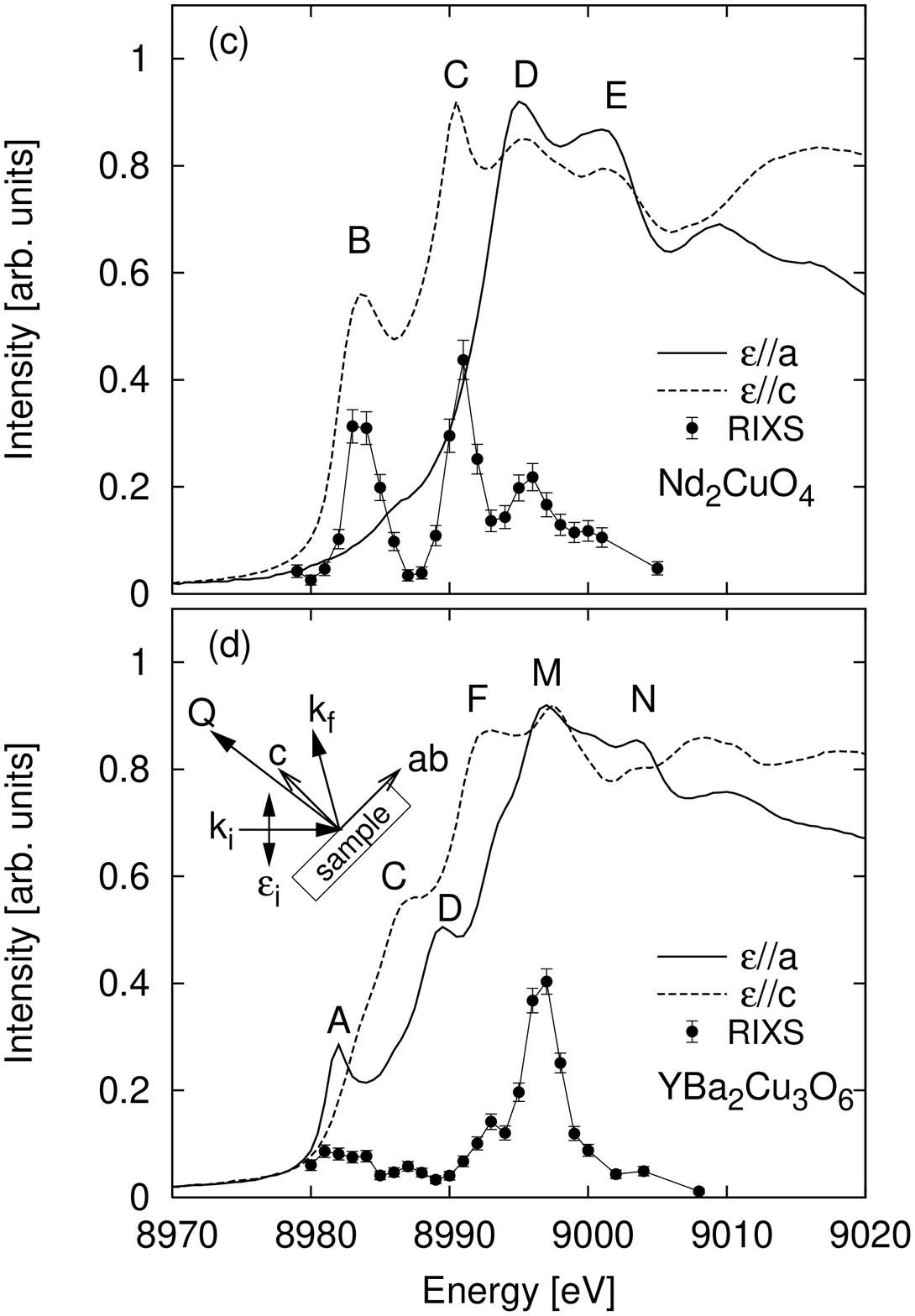}
\caption{Incident photon energy dependence of RIXS spectra of (a)
Nd$_2$CuO$_4$ and (b) YBa$_2$Cu$_3$O$_6$.  The absolute momentum
transfer is $\vec{Q}=(0.5,0,12.5)$ and $(0.5,0,12.3)$ for Nd$_2$CuO$_4$
and YBa$_2$Cu$_3$O$_6$, respectively. The incident energy for each scan
can be read from the vertical axis.  X-ray absorption spectra of (c)
Nd$_2$CuO$_4$ and (d) YBa$_2$Cu$_3$O$_6$.  RIXS intensities integrated
between 1 and 3 eV are also shown as a function of the incident photon
energy.}
\label{fig:eidep}
\end{figure*}


Figures \ref{fig:eidep}(a) and (b) show the incident energy dependence
of the RIXS spectra of Nd$_2$CuO$_4$ and YBa$_2$Cu$_3$O$_6$,
respectively.  The reduced momentum transfer in the $ab$ plane
($\vec{q}$) is $(\pi,0)$.  The spectral weight around 2 eV arises from
the excitation across the Mott gap and its intensity is resonantly
enhanced at some incident energies.

Because the intermediate state of RIXS corresponds to the final state of
x-ray absorption, it is interesting to compare absorption spectra with
the incident photon energy dependence of RIXS.  Solid and dashed lines
in Figs.\ \ref{fig:eidep}(c) and (d) are the absorption spectra of
Nd$_2$CuO$_4$ and YBa$_2$Cu$_3$O$_6$ obtained by the total fluorescence
yield.  Prominent peaks in the absorption spectra are labeled by the
same letters A-N as in the reports \cite{Kosugi1990,Tolentino1992}.  In
Nd$_2$CuO$_4$, the peaks B and C in the $\epsilon \parallel c$ spectrum
are assigned to the $1s$-$4p_{\pi}$ transitions.  The peaks D and E in
the $\epsilon \parallel a$ spectrum are the $1s$-$4p_{\sigma}$
transitions. In each pair, the peaks at lower energy (B and D) and those
at higher energy (C and E) correspond to well-screened and
poorly-screened core-hole final states, respectively \cite{Kosugi1990}.
On the other hand, the assignment of the features in the
YBa$_2$Cu$_3$O$_6$ spectra is rather complicated \cite{Tolentino1992}
because there are two distinct Cu sites.  One type of Cu atoms is in the
monovalent Cu(1) site, which forms the one-dimensional chains in
YBa$_2$Cu$_3$O$_7$.  The other is the divalent Cu(2) site in the CuO$_2$
plane.  The peaks of A and D in the $\epsilon \parallel a$ spectrum are
absent in YBa$_2$Cu$_3$O$_{6.95}$ and they are attributed to the
transitions at the Cu(1) sites.  The peak A is a proof of the existence
of a monovalent Cu atom \cite{Heald1988,Tranquada1988}.  The peaks M and
N are the $1s$-$4p_{\sigma}$ transitions of the Cu(2) site.  A pair of
peaks of the $1s$-$4p_{\pi}$ transitions of the Cu(2) site should appear
in the $\epsilon \parallel c$ spectrum.  It can be assigned to the peaks
C and F but the $1s$-$4p_{\sigma}$ transition of the Cu(1) site may
overlap.

In YBa$_2$Cu$_3$O$_6$, a large RIXS intensity is observed around 5 eV
when the incident photon energy is close to the peak A in the absorption
spectra, indicating that this excitation is related to the Cu(1) site.
In order to show the resonant feature of the Mott gap excitation more
precisely, the RIXS intensity in Fig.\ \ref{fig:eidep} (a) and (b) is
integrated between 1 and 3 eV and plotted as a function of the incident
photon energy in Figs.\ \ref{fig:eidep}(c) and (d).  Clear difference is
observed in the incident energy profile of the Mott gap excitation
between Nd$_2$CuO$_4$ and YBa$_2$Cu$_3$O$_6$, that is, there are three
well-pronounced resonances for the former and only one peak with a bump
on the lower-energy side for the latter.  Each resonant energy of the
RIXS intensity is found to correspond to a peak in the absorption
spectra.  We note that both out-of-plane ($\vec{\epsilon_i} \parallel
\vec{c}$) and in-plane ($\vec{\epsilon_i} \perp \vec{c}$) components
coexist in the incident photon polarization in our experimental
condition.  In contrast to the two cases, La$_2$CuO$_4$ shows two
distinct resonances with respect to the polarization condition; one is
at $E_i$ = 8990 eV in the out-of-plane condition and the other is at
$E_i$ = 8994 eV in the in-plane condition \cite{Lu2006}.


As shown in Fig.\ \ref{fig:rixs}, the screening process by the charge
transfer from the ligand oxygen to the UHB in the intermediate state is
important for the Mott gap excitation.  This means that the Mott gap
excitation is enhanced when the incident energy is tuned to the
well-screened state in the x-ray absorption spectra
\cite{Id'e1999,Id'e2000}.  This rule is applicable to La$_2$CuO$_4$
\cite{Lu2006} as well as to the in-plane condition of Nd$_2$CuO$_4$
[peak D in Fig.\ \ref{fig:eidep}(c)] and YBa$_2$Cu$_3$O$_6$ [peak M in
Fig.\ \ref{fig:eidep}(d)].  In contrast, two resonances appear in the
out-of-plane condition of Nd$_2$CuO$_4$ [peak B and C in Fig.\
\ref{fig:eidep}(c)].  In the configuration-interaction picture, the
well-screened and poorly-screened states are expressed by linear
combinations of $\vert\underline{1s},3d^9,4p_{\pi}\rangle$ and
$\vert\underline{1s},3d^{10},\underline{L},4p_{\pi}\rangle$, where
$\underline{1s}$ and $\underline{L}$ denote a hole in $1s$ level and
ligand oxygen, respectively.  Generally,
$\vert\underline{1s},3d^{10},\underline{L},4p_{\pi}\rangle$ is dominant
in the well-screened state.  The two resonances in the out-of-plane
condition of Nd$_2$CuO$_4$ indicate that
$\vert\underline{1s},3d^{10},\underline{L},4p_{\pi}\rangle$ has
relatively large weight at the absorption labeled C in Fig.\
\ref{fig:eidep}(c), though it has been assigned to the poorly-screened
state.  We have no explanation for the weakness or disappearance of the
resonance in the out-of-plane condition of YBa$_2$Cu$_3$O$_6$ so far.
It may be related to a structural difference, namely, the Cu atom in the
CuO$_2$ plane is fivefold coordinated by four planar and one apical
oxygens in YBa$_2$Cu$_3$O$_6$, which breaks the mirror symmetry
perpendicular to the c-axis, while the symmetry is present in
La$_2$CuO$_4$ and Nd$_2$CuO$_4$.  Further systematic studies are
necessary to understand the difference in the out-of-plane condition.

\subsection{Momentum dependence of the Mott insulators}
\begin{figure*}[phtb]
\includegraphics[scale=0.34]{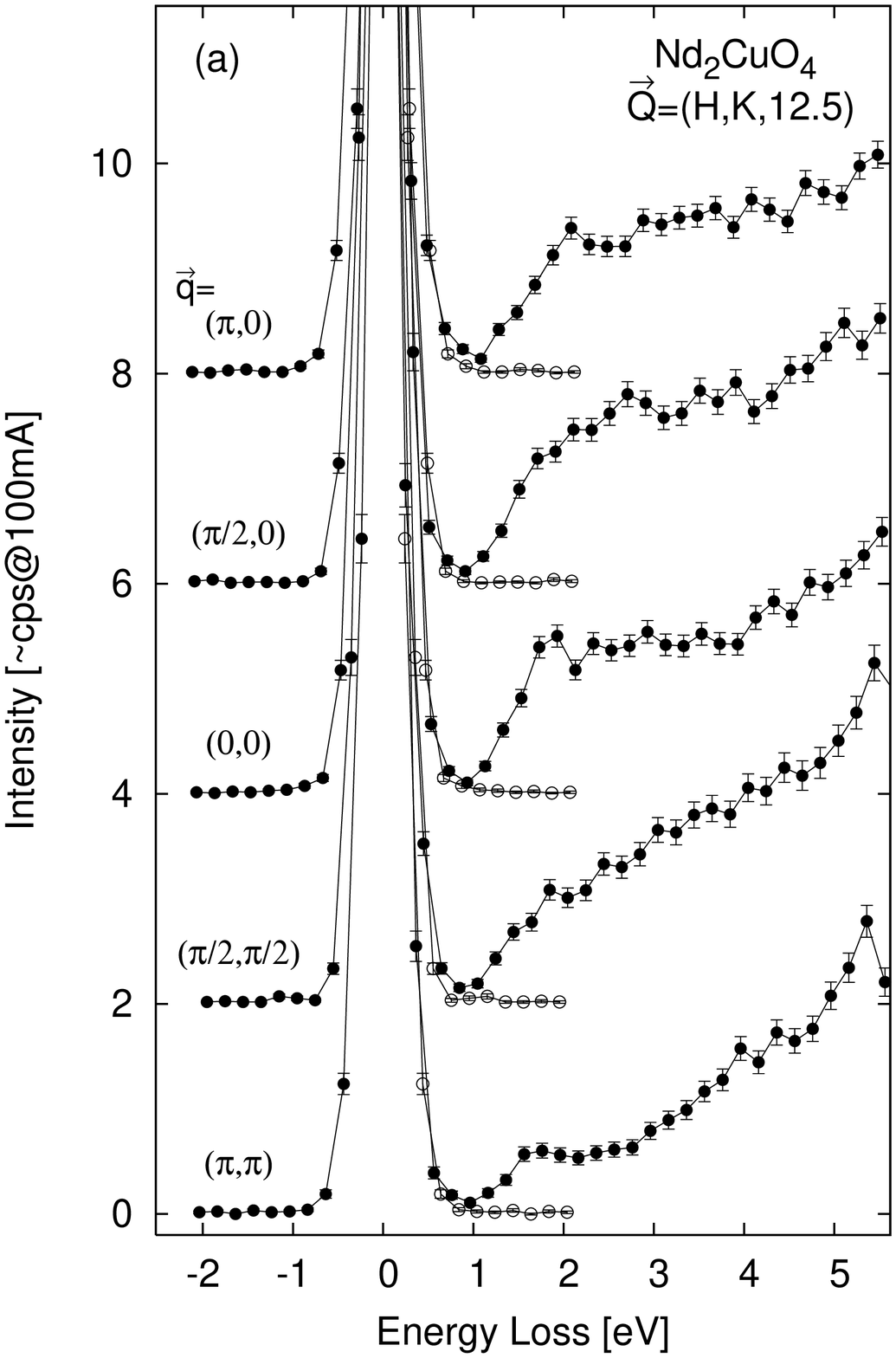}
\includegraphics[scale=0.34]{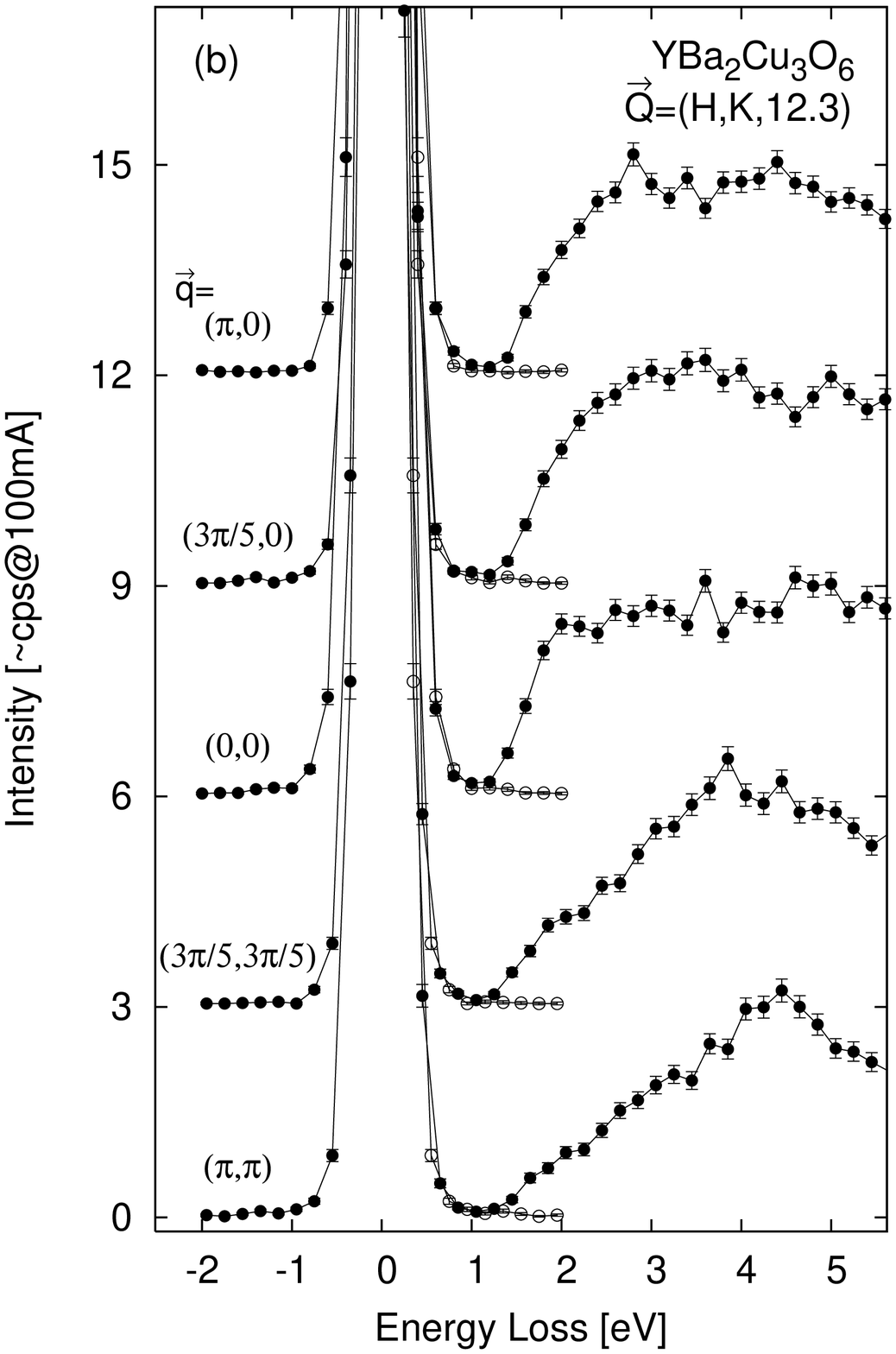}
\includegraphics[scale=0.34]{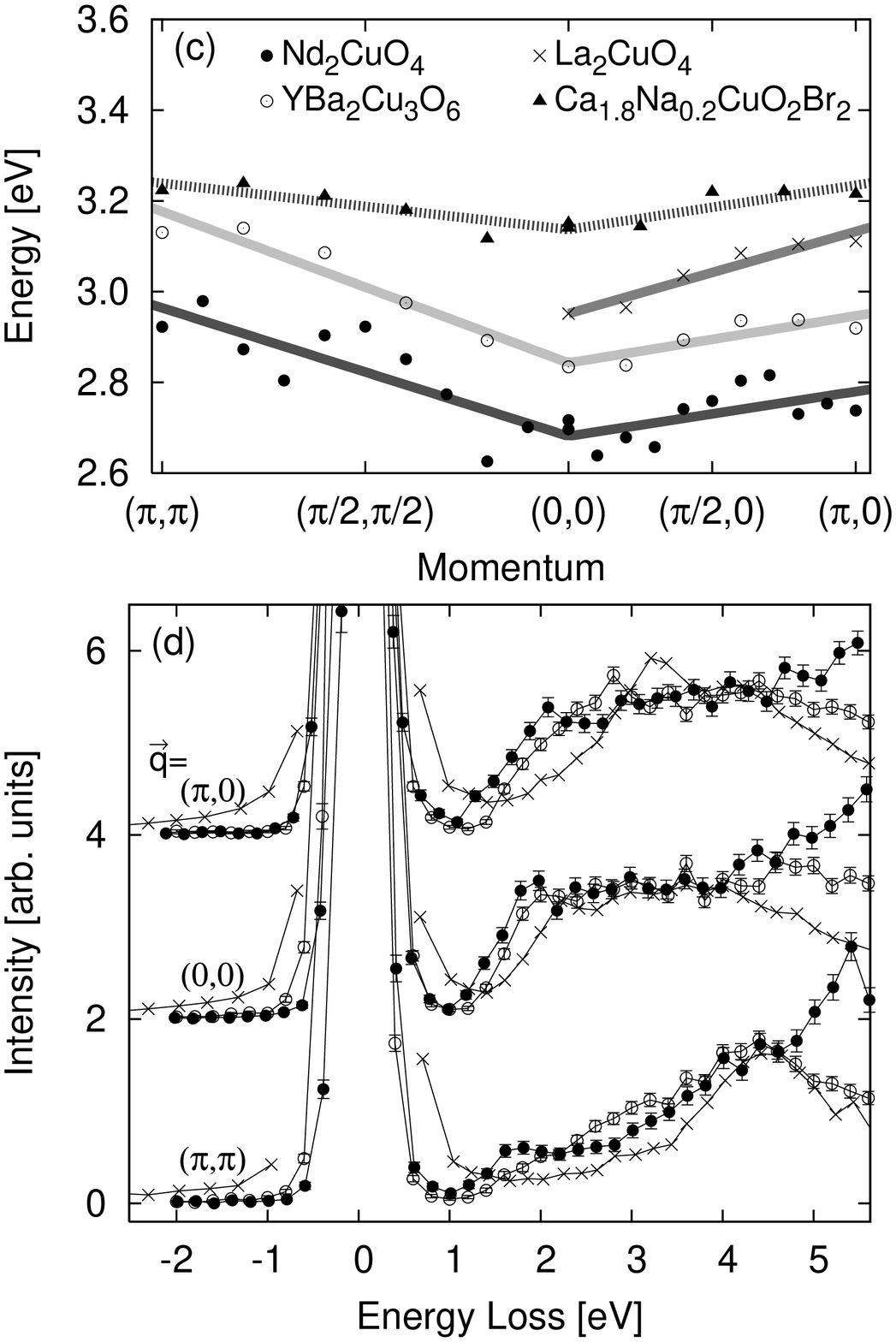} \caption{Momentum
dependence of RIXS spectra of (a) Nd$_2$CuO$_4$, (b) YBa$_2$Cu$_3$O$_6$.
The incident photon energy is 8991 eV for Nd$_2$CuO$_4$ and 8996 eV for
YBa$_2$Cu$_3$O$_6$.  Filled circles are the raw data.  The data in the
anti-Stokes region are folded at the origin and plotted as open circles,
from which we can estimate the elastic tail and the background.  (c)
Momentum dependence of the Mott gap excitation.  The center of gravity
of the RIXS spectra between 1 and 4 eV is plotted for Nd$_2$CuO$_4$,
YBa$_2$Cu$_3$O$_6$, and La$_2$CuO$_4$.  The data of La$_2$CuO$_4$ are
taken from the Ref.\ \cite{Collart2006}.  For
Ca$_{1.8}$Na$_{0.2}$CuO$_2$Br$_2$, the center of gravity between 2 and 4
eV is used to exclude contribution from the intraband excitation.  The
thick lines are drawn assuming a linear dispersion.  (d) Comparison of
three Mott insulators. The symbols are the same as in (c).  The data of
La$_2$CuO$_4$ are taken from the Ref.\ \cite{Kim2002}. }
\label{fig:qdep}
\end{figure*}

Figures \ref{fig:qdep} (a) and (b) show the momentum dependence of
Nd$_2$CuO$_4$ and YBa$_2$CuO$_6$, respectively.  The incident photon is
fixed at the resonance energy derived from Figs.\ \ref{fig:eidep}.  The
RIXS intensity is almost zero at $\sim$ 1 eV and a clear Mott gap can be
seen at all momenta.  Taking advantage of the experimental condition of
$2\theta \simeq 90^{\circ}$ , the gap is much clearer in our spectra
than that in the previous RIXS works on parent materials,
Ca$_2$CuO$_2$Cl$_2$ \cite{Hasan2000a} and La$_2$CuO$_4$ \cite{Kim2002}
where the scattering intensity below the gap does not reach close to
zero due to the tail of the elastic scattering.

Here, we discuss the momentum dependence of the overall spectral weight
rather than that of the peak positions.  A recent detailed study of
HgBa$_2$CuO$_{4+\delta}$ and La$_2$CuO$_4$ demonstrated that a multiplet
of peaks with a weak dispersion can be elucidated by utilizing the
subtle dependence of the cross section on the incident photon energy
\cite{Lu2005,Lu2006}.  Though the peak position of each peak shows a
weak dispersion, the overall spectral weight containing the intensity of
the multiplet shifts as a function of momentum transfer.  Moreover, such
a multiplet is not clearly observed in our data, which may be due to
different experimental conditions with Ref.\ \cite{Lu2005,Lu2006}, such
as polarization and absolute momentum transfer.  Focusing on the
spectral weight, we notice from Figs.\ \ref{fig:qdep} (a) and (b) that
the changes in the excitation spectra are small along the $[\pi,0]$
direction while the spectral weight clearly shifts to higher energy as
the momentum transfer increases along the $[\pi,\pi]$ direction.
The center of gravity of the RIXS spectra between 1 and 4 eV
is plotted as a function of momentum transfer in Fig.\
\ref{fig:qdep}(c), which corresponds to the dispersion relation of the
spectral weight of the Mott gap excitation.  The momentum dependence of
the Mott gap excitation shows a large anisotropy; the dispersion along the
$[\pi,\pi]$ direction is larger than that along the $[\pi,0]$ direction.

The larger dispersion along the $[\pi,\pi]$ direction was predicted by a
theoretical calculation based on a single band Hubbard model with long
range hoppings ($t$-$t'$-$t''$-$U$ model) \cite{Tsutsui1999}. Besides,
it has been experimentally observed in Ca$_2$CuO$_2$Cl$_2$
\cite{Hasan2000a}.  In La$_2$CuO$_4$ \cite{Kim2002}, the authors claimed
that the spectral weight of the lowest excitation (labeled A in the
reference) becomes small at $\vec{q}=(\pi,\pi)$ and this results in a
shift of the spectral weight to higher energy.  In Fig.\
\ref{fig:qdep}(d), we compare the RIXS spectra of La$_2$CuO$_4$ in Ref.\
\cite{Kim2002} with our data. The momentum dependence of La$_2$CuO$_4$
is qualitatively similar to that of Nd$_2$CuO$_4$ and
YBa$_2$Cu$_3$O$_6$.  Therefore, we conclude that the larger dispersion
along the $[\pi,\pi]$ direction than that along the $[\pi,0]$ direction
is a common character of the Mott gap excitation in the insulating
CuO$_2$ plane.

Quantitatively, the RIXS spectra depend on the material.  Comparing the
spectra at the zone center in Fig.\ \ref{fig:qdep}(d), the gap energy is
found to systematically change with the number of oxygens coordinated
around Cu, that is, it is largest in the octahedral structure of
La$_2$CuO$_4$ and smallest in the square structure of Nd$_2$CuO$_4$.
YBa$_2$Cu$_3$O$_6$, with its pyramidal coordination geometry, lies
between the two.  This systematic change of the gap has been already
reported from an optical conductivity experiment \cite{Tokura1990},
which is consistent with our RIXS results.

The magnitude of the dispersion also displays material dependence.
Experimentally, the dispersion along the $[\pi,0]$ direction is apparent
in La$_2$CuO$_4$ \cite{Kim2002,Lu2006,Collart2006}, while that of
Nd$_2$CuO$_4$ and YBa$_2$Cu$_3$O$_6$ is fairly small.  The dispersion of
La$_2$CuO$_4$ in Ref.\ \cite{Collart2006}, where momentum-dependent RIXS
spectra along the $[\pi,0]$ direction are presented, is superimposed in
Fig.\ \ref{fig:qdep}(c). We can also ascertain from the spectral weight
that the dispersion along $[\pi,0]$ direction is larger in
La$_2$CuO$_4$.  On the other hand, it is demonstrated theoretically that
the long-range hoppings $t'$ and $t''$ play an important role for the
dispersion of RIXS spectra \cite{Tsutsui1999,Tohyama2005}; the
dispersion along the $[\pi,0]$ direction becomes larger without $t'$ and
$t''$.  Accordingly, our experimental results suggest that long range
hopping parameters of Nd$_2$CuO$_4$ and YBa$_2$Cu$_3$O$_6$ are larger
than those of La$_2$CuO$_4$.  Comparing relatively between Nd$_2$CuO$_4$
and La$_2$CuO$_4$, it is consistent with the parameters which are
obtained from the shape of the Fermi surface in ARPES experiments
\cite{Tohyama2000}.  Furthermore the experimental result that
YBa$_2$Cu$_3$O$_6$ has larger long range hopping than La$_2$CuO$_4$
agrees with a theory which proposed that materials with higher $T_c$ at
optimal doping have larger hopping term \cite{Pavarini2001}.


\subsection{Carrier-doping effect}
\begin{figure*}[phtb]
\includegraphics[scale=0.34]{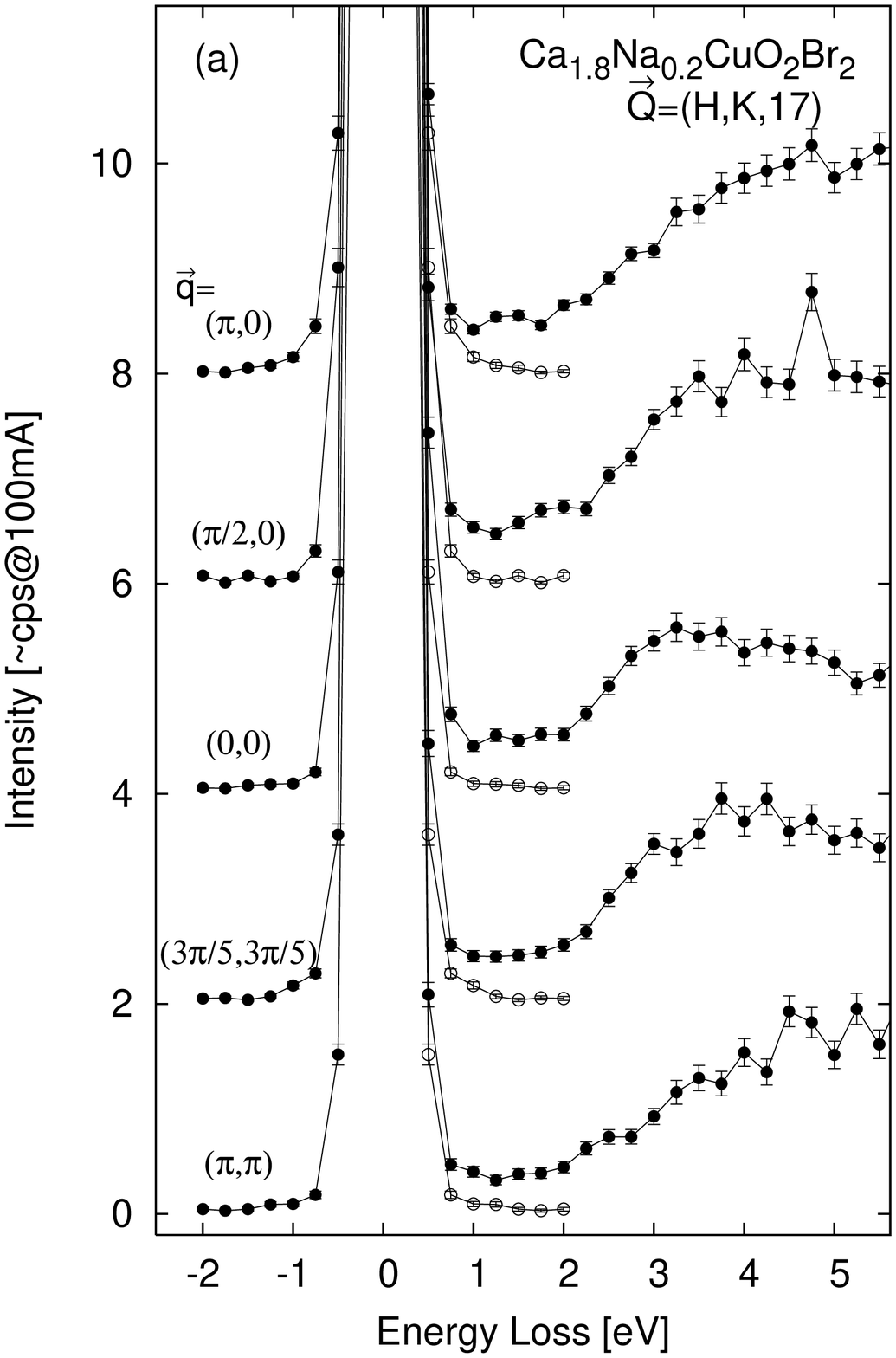}
\includegraphics[scale=0.34]{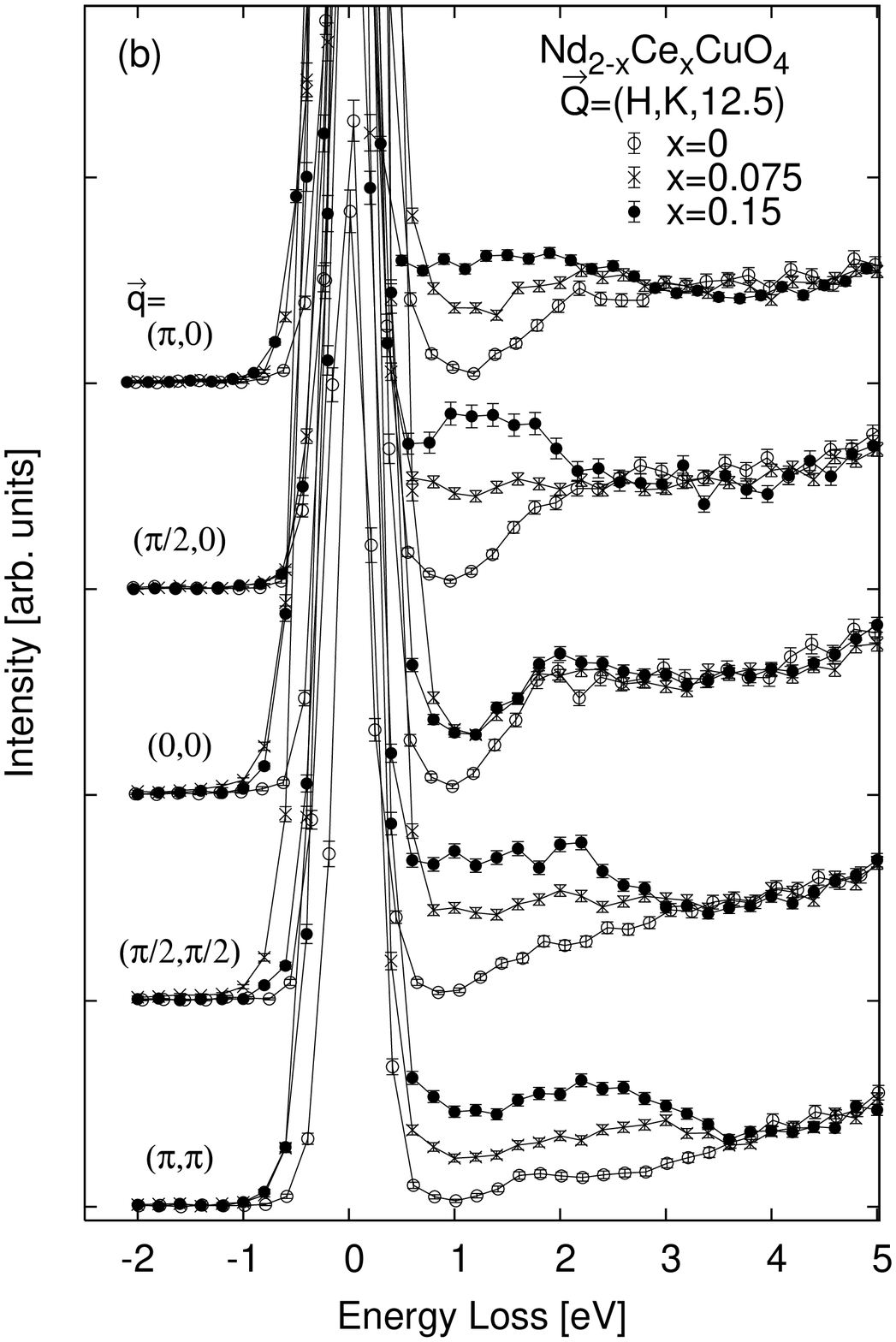}
\includegraphics[scale=0.34]{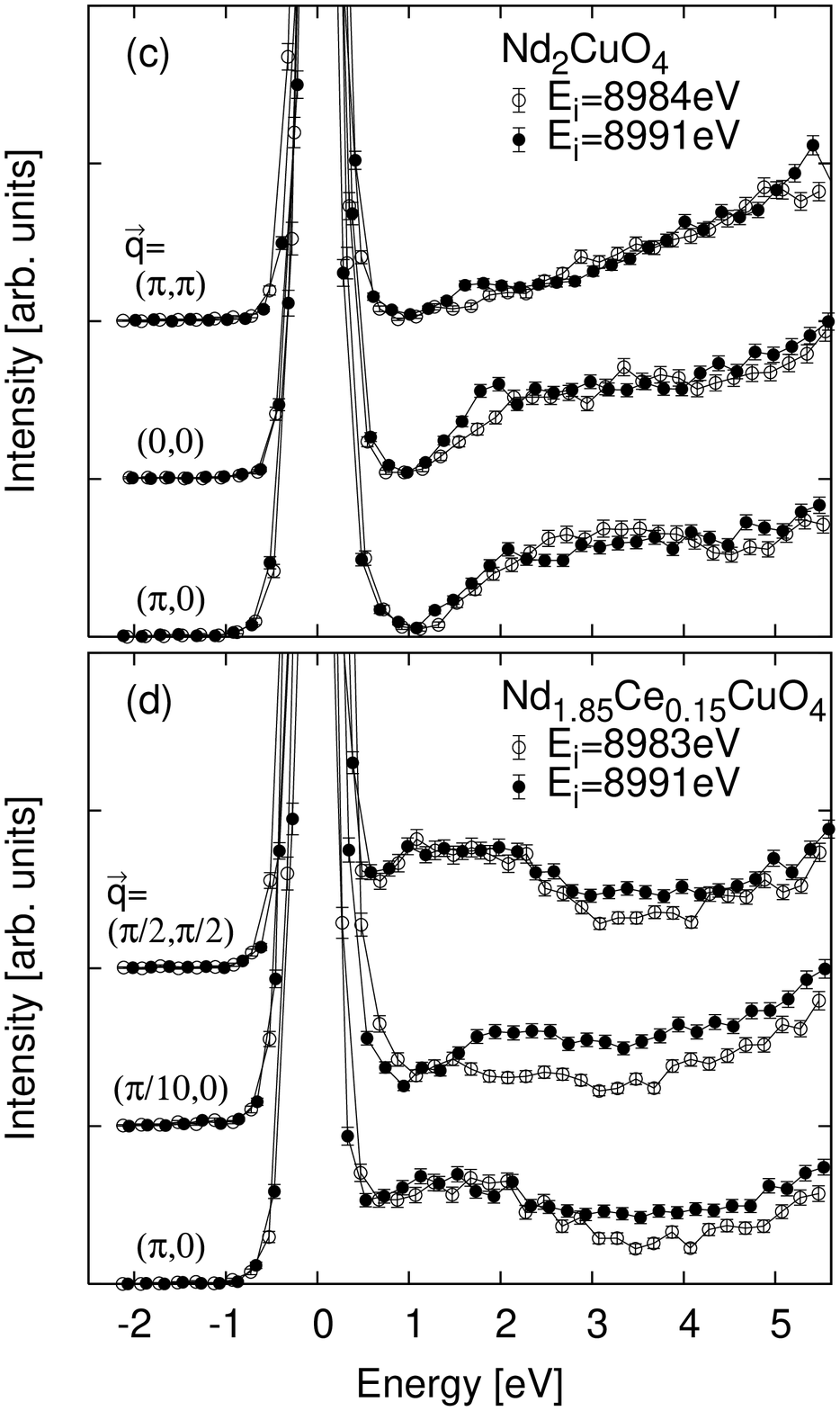}
\caption{(a) RIXS spectra of hole-doped
Ca$_{1.8}$Na$_{0.2}$CuO$_2$Br$_2$.  The incident photon energy is 8984
eV.  Filled circles are the raw data.  The data in the anti-Stokes
region are folded at the origin and plotted as open circles.  (b) RIXS
spectra of Nd$_{2-x}$Ce$_x$CuO$_4$, which show electron-doping
dependence.  The incident photon energy is 8991 eV.  (c) and (d)
Comparison of RIXS spectra of Nd$_2$CuO$_4$ and
Nd$_{1.85}$Ce$_{0.15}$CuO$_4$ measured at two incident photon energies.}
\label{fig:doped}
\end{figure*}

In this section, we discuss RIXS spectra of carrier-doped
superconductors and compare them with those of Mott insulators.  Fig.\
\ref{fig:doped}(a) shows the momentum dependence of RIXS spectra
obtained for hole-doped Ca$_{1.8}$Na$_{0.2}$CuO$_2$Br$_2$.  The incident
photon energy is 8984 eV which corresponds to the well-screened state of
the out-of-plane condition in the absorption spectrum.  A large spectral
weight remains located above 2 eV, which indicates that the Mott gap
persists even in the hole-doped superconductor.  At the same time, the
gap is filled by a continuum intensity below 2 eV.  While the former
excitation is the interband excitation across the Mott gap, the latter
is related to the dynamics of doped holes in the Zhang-Rice singlet band
and we call it intraband excitation.

When the interband excitation of Ca$_{1.8}$Na$_{0.2}$CuO$_2$Br$_2$ is
compared with that of Mott insulators, the spectral weight around 2 eV
decreases by hole doping except for the spectrum at $\vec{q}=(\pi,\pi)$
where the intensity around 2 eV is weak even in Mott insulators.  As a
result, the total spectral weight of the interband excitation shifts to
higher energy and the dispersion becomes weaker.  In Fig.\
\ref{fig:qdep}(c), the center of gravity of the spectral weight between 2-4
eV is shown as a function of momentum.  It is apparent that the
dispersion is weaker in the hole-doped Ca$_{1.8}$Na$_{0.2}$CuO$_2$Br$_2$
than in the Mott insulators.  Such a change is also observed in
La$_{2-x}$Sr$_x$CuO$_4$ \cite{Kim2004c}; a well-defined peak feature at
2-3 eV for $x$ = 0 (labeled A in Ref.\ \cite{Kim2002}) decreases in
intensity or disappears at $x$ = 0.17 and the overall spectral weight (labeled
AB in Ref.\ \cite{Kim2004c}) shows weaker dispersion.  Furthermore, the
excitation from the CuO$_2$ plane in optimally-doped
YBa$_2$Cu$_3$O$_{7-\delta}$ also shows very weak dispersion
\cite{Ishii2005a,Ishii2006a}.

These characteristic changes of the Mott gap excitation by hole doping
are consistent with a theoretical prediction
\cite{Tsutsui2003,Tohyama2005}.  The shift of the spectral weight to
higher energy is attributed to a shift of the Fermi energy upon hole
doping.  On the other hand, it is suggested in the theory that the
weaker dispersion in the hole-doped case compared with that in the
undoped case is related to the reduction of antiferromagnetic
correlation.  With realistic parameters used in the theory, the
antiferromagnetic correlation is strongly suppressed by hole doping and
the dispersion of the Mott gap excitation becomes weaker.  In contrast,
the short-range antiferromagnetic correlation is kept in the
electron-doped case, and the magnitude of the dispersion of the Mott gap
excitation is almost the same as that of the undoped case.  This result
indicates that the underlying magnetism governs the charge dynamics in
the high-energy region up to a few eV.  It is noted that the two-leg
ladder shows a different behavior upon hole doping.  A recent RIXS study
of (La,Sr,Ca)$_{14}$Cu$_{24}$O$_{41}$ demonstrated that the momentum
dependence of the Mott gap excitation in the two-leg ladder is
insensitive to the hole doping as well as the spin gap state
\cite{Ishii2007}.

Next, we discuss the results of electron doping.  In Fig.\
\ref{fig:doped}(b), RIXS spectra for the $x$ = 0, 0.075, and 0.15 in
samples of Nd$_{2-x}$Ce$_x$CuO$_4$ are compared.  The incident photon
energy is fixed at 8991 eV for all samples.  The spectra are normalized
to the intensity at 4-5 eV.  The doping dependence is apparently
different between the zone center and finite $\vec{q}$.  At the zone
center, the Mott gap excitation at 2 eV is unchanged upon electron
doping, which means that the charge gap due to the electron correlation
persists even in the metallic phase.

Two theoretical calculations of the electron doped CuO$_2$ plane have
been reported.  One is based on a single band Hubbard model using a
numerically exact diagonalization technique on a small cluster
\cite{Tsutsui2003}.  The other adopts a $d$-$p$ model and the
Hartree-Fock approximation is applied \cite{Markiewicz2006}.  In the
former theory, the Mott gap excitation persists upon electron doping,
and especially, it is enhanced in the spectrum at the zone center.
On the other hand, the latter suggests a collapse of the gap.  Our
experimental observation of the 2 eV feature remaining in the electron-doped
metallic state supports the validity of the former theory.

At finite momentum transfer, large spectral weight appears below the
gap.  Its intensity is roughly proportional to the number of doped
electrons ($x$).  Hence it is related to the intraband dynamics of the
doped electrons in the UHB.  Because details of the intraband excitation
have already been published \cite{Ishii2005b,Ishii2006}, we do not
discuss them here.

Finally, we show the difference of resonance conditions between the
undoped and electron-doped cases.  In Fig.\ \ref{fig:doped}(c), RIXS
spectra measured at two incident photon energies are compared for
Nd$_2$CuO$_4$.  The lower 8984 eV is close to the well-screened state of
the out-of-plane condition [peak B in Fig.\ \ref{fig:eidep}(c)], while
the higher 8991 eV corresponds to the poorly-screened state [peak C in
Fig.\ \ref{fig:eidep}(c)].  The scale factors are common for the spectra
at the three momenta.  The spectra strongly resemble between the two
incident photon energies.  On the other hand, the spectra of
Nd$_{1.85}$Ce$_{0.15}$CuO$_4$ in Fig.\ \ref{fig:doped}(d) show a
difference near the zone center, $\vec{q}=(\pi/10,0)$; the 2 eV feature
is suppressed at 8983 eV.  This is an additional proof that the
excitation at 2 eV at the zone center is qualitatively different from
the intraband excitation observed at the finite momentum transfer.  In
the electron-doped case, only the intraband excitation is expected to be
enhanced at the incident photon energy where the core hole is created at
the doped site \cite{Ishii2005b,Tohyama2005}.  This energy should exists
below the well-screened state and 8983 eV in
Nd$_{1.85}$Ce$_{0.15}$CuO$_4$ may be close to it.

\section{Summary}
We have performed a Cu $K$-edge RIXS study of high-$T_c$ cuprates from
parent Mott insulators to carrier-doped superconductors.  While the
resonance condition depends on materials, an anisotropic dispersion,
that is, larger dispersion along the $[\pi,\pi]$ direction than the
$[\pi,0]$ direction, is commonly observed in undoped Mott insulators.
Upon hole doping, the dispersion of the Mott gap excitation becomes
weaker, while the Mott gap excitation is prominent at the zone center in
the electron doped case.  At the same time, an intraband excitation
emerges below the gap in both hole- and electron-dopings.  These
characteristics are consistent with a theoretical calculation based on
the Hubbard model.

\section*{Acknowledgments}
The authors thank Dr.\ I.\ Jarrige for proofreading of the manuscript.
This work was performed under the inter-university cooperative research
program of the Institute of Materials Research, Tohoku University and
financially supported by the Grant-in-Aid for Scientific Research on
Priority Areas "Invention of anomalous quantum materials" from the
Ministry of Education, Culture, Sports, Science, and Technology of
Japan.  K.\ I. was also supported by Grant-in-Aid for Young Scientists
from JSPS.  The crystal growth of YBa$_2$Cu$_3$O$_6$ was supported by
the New Energy and Industrial Technology Development Organization (NEDO)
as the Collaborative Research and Development of Fundamental
Technologies for Superconductivity Applications.

\bibliographystyle{elsart-num}
\bibliography{/home/kenji/tex/jabref/bib/high-tc.bib,/home/kenji/tex/jabref/bib/rixs-experiment.bib,/home/kenji/tex/jabref/bib/ixs-theory.bib}

\end{document}